\documentclass[fleqn,twoside]{article}
\usepackage{espcrc2}
\usepackage{graphicx}
\usepackage[figuresright]{rotating}

\newcommand{\AmS}{{\protect\the\textfont2
  A\kern-.1667em\lower.5ex\hbox{M}\kern-.125emS}}

\def\Journal#1#2#3#4{{#1} {\bf #2}, #3 (#4)}

\def\AP{\em Annals of Phys.}

\def\EPJC{{\em Eur. Phys. Jour.} C}
\def\JHEP{\em Jour. of High Energy Phys.}

\def\NPB{{\em Nucl. Phys.} B}
\def\PA{{\em Physica A}}
\def\PLB{{\em Phys. Lett.}  B}

\def\PRD{{\em Phys. Rev.} D}

\newcommand{\ket}{\,\rangle}
\newcommand{\bra}{\langle \,}

\def\p{\pi}
\def\t{\tau}
\def\m{\mu}
\def\n{\nu}

\title{Improving the Hadronization of QCD currents in TAUOLA and PHOKHARA}

\author{P. Roig\address{Instituto de F\'isica Corpuscular, IFIC, CSIC-Universitat de Val\`encia.\\ Apt. de Correus 22085, E-46071 Val\`encia, Spain}%
        \thanks{I thank Stephan Narison and his collaborators for the organization of the QCD08 Conference. This work has been supported in part by the EU MRTN-CT-2006-035482 (FLAVIAnet), by MEC (Spain) under grant FPA2007-60323 and by the Spanish Consolider-Ingenio 2010 Program CPAN (CSD2007-00042).}}
\begin{document}
\thispagestyle{empty}
\begin{abstract}
\noindent We present our study of the hadronization structure of both vector and axial-vector currents leading to decays of the tau into two kaons and a pion. The cornerstones of our framework are the large-$N_C$ limit of QCD, the chiral structure exhibited at low energies and the proper asymptotic behaviour, ruled by QCD, that is demanded to the associated form factors. The couplings of the theory are mostly constrained by this procedure and upon the analysis of BaBar data on $e^+e^-\to KK\pi$ we are able to predict the hadronic spectra.
\end{abstract}

\maketitle

\section{INTRODUCTION} \label{intro}
\noindent We focus on the nonperturbative dynamics of the strong interaction that can be learned thanks to the clean separation of the electroweak and QCD parts in the hadronic decays of the $\t$ lepton. Advances in the study of the two meson processes have been presented in this workshop \cite{Guo},\cite{Jorge}.\\
\noindent The decay amplitude for the considered decays may be written as:
\begin{equation} \label{Mgraltau}
\mathcal{M}\,=\,-\frac{G_F}{\sqrt{2}}\,V_{\mathrm{ud/us}}\,\overline{u}_{\nu_\tau}\gamma^\mu(1-\gamma_5)\,u_\tau \mathcal{H}_\mu\,,
\end{equation}
where our lack of knowledge of the precise hadronization mechanism is encoded in the hadronic vector, $\mathcal{H}_\mu$:
\begin{equation} \label{Hmugral}
\mathcal{H}_\mu = \bra \left\lbrace  P(p_i)\right\rbrace_{i=1}^n |\left( \mathcal{V}_\mu - \mathcal{A}_\mu\right)  e^{i\mathcal{L}_{QCD}}|0\ket\,.
\end{equation}
\noindent Symmetries help us to decompose $\mathcal{H}_\mu$ depending on the number of final-state pseudoscalar ($P$) mesons, $n$. For three mesons in the final state, this reads:
\begin{eqnarray} \label{Hmu3m}
\mathcal{H}_\mu = V_{1\mu} F_1^A(Q^2,s_1,s_2) + V_{2\mu} F_2^A(Q^2,s_1,s_2) +\nonumber\\
 Q_\mu F_3^A(Q^2,s_1,s_2) + i \,V_{3\mu} F_4^V(Q^2,s_1,s_2)\,,
\end{eqnarray}
and
\begin{eqnarray} \label{VmuQmu}
V_{1\mu} &  = & \left( g_{\mu\nu} - \frac{Q_{\mu}Q_{\nu}}{Q^2}\right) \,
(p_2 - p_1)^{\nu} ,\nonumber\\
V_{2\mu} & = & \left( g_{\mu\nu} - \frac{Q_{\mu}Q_{\nu}}{Q^2}\right) \,
(p_3 - p_1)^{\nu}\,,\nonumber\\
V_{3\mu} & = & \varepsilon_{\mu\nu\varrho\sigma}\,p_1^\nu\, \,p_2^\varrho\, \,p_3^{\sigma} ,\nonumber\\ Q_\mu & = & (p_1\,+\,p_2\,+\,p_3)_\mu \,,\,s_i = (Q-p_i)^2\,.
\end{eqnarray}
\noindent $F_i$, $i=1,2,3$, correspond to the axial-vector current ($\mathcal{A}_\m$) while $F_4$ drives the vector current ($\mathcal{V}_\m$). The form factors $F_1$ and $F_2$ have a transverse structure in the total hadron momenta, $Q^\mu$, and drive a $J^P=1^+$ transition. The scalar form factor, $F_3$, vanishes with the mass of the Goldstone bosons (chiral limit) and, accordingly, gives a tiny contribution. This is as far as we can go without model assumptions, that is, we are still not able to derive the $F_i$ from QCD. Our claim is that the most adequate way out is the use of a phenomenologically motivated theory that, in the energy region spanned by tau decays, resembles QCD as much as possible.\\
\noindent In any case, the approximate symmetries of QCD are useful. They rule what is the theory to be used in its very low-energy domain and guide the construction of higher-energy theories
.
\section{STRONG INTERACTION AMONG LIGHT-FLAVOURED MESONS} \label{QCDMesons}
\noindent Weinberg's Theorem \cite{wei} leads to conclude that a description in terms of the relevant degrees of freedom in the range of energies spanned by tau decays that incorporates the (approximate) symmetries of QCD therein will be valid. Some phenomenological wisdom would tell that this procedure will be most adequate, indeed. Well below the $\rho$(770) mass, $\chi$PT \cite{wei}, \cite{cpt} is the effective theory of QCD. Being $M_\tau\sim1.8$ GeV, one needs to include the resonances as explicit degrees of freedom. According to vector meson dominance \cite{vmd}, those of spin one would prevail. This is a result derived within Resonance Chiral Theory (R$\chi$T) \cite{rcht1} worked in the large-$N_C$ limit of QCD. R$\chi$T is built upon the approximate chiral symmetry of low-energy QCD for the lightest pseudoscalar mesons and unitary symmetry for the resonances. The perturbative expansion of R$\chi$T is guided by its large-$N_C$ limit \cite{lnc}. At LO in 1/$N_C$, meson dynamics is described by tree level diagrams obtained from an effective local Lagrangian including the interactions among an infinite number of stable resonances. We include finite resonance widths (a NLO effect in this expansion) within our framework \cite{width}. We are also departing from the $N_C\to\infty$ limit because we consider just one multiplet of resonances per set of quantum numbers 
 and not the infinite tower predicted that cannot be included in a model independent way. For convenience, we choose to represent the spin-one mesons in the antisymmetric tensor formalism \cite{rcht2}.\\
\noindent The relevant part of the R$\chi$T Lagrangian is \cite{rcht1}, \cite{vap}, \cite{vvp}, \cite{paper}:
\begin{eqnarray} \label{Full_Lagrangian}
& & \mathcal{L}_{R\chi T}=\frac{F^2}{4}\bra u_\mu u^\mu +\chi_+ \ket+\frac{F_V}{2\sqrt{2}}\bra V_{\mu\nu} f^{\mu\nu}_+ \ket \nonumber\\
& & + \frac{i \,G_V}{\sqrt{2}} \bra V_{\mu\nu} u^\mu u^\nu\ket + \frac{F_A}{2\sqrt{2}}\bra A_{\mu\nu} f^{\mu\nu}_- \ket +\mathcal{L}_{\mathrm{kin}}^V\nonumber\\
& & + \mathcal{L}_{\mathrm{kin}}^A + \sum_{i=1}^{5}\lambda_i\mathcal{O}^i_{VAP} + \sum_{i=1}^7\frac{c_i}{M_V}\mathcal{O}_{VJP}^i \nonumber\\
& & + \sum_{i=1}^4d_i\mathcal{O}_{VVP}^i + \sum_{i=1}^5 \frac{g_i}{M_V} {\mathcal O}^i_{VPPP} \, ,
\end{eqnarray}
where all couplings are real, being $F$ the pion decay constant in the chiral limit. The notation is that of Ref.~\cite{rcht1}. $P$ stands for the lightest pseudoscalar mesons and $A$ and $V$ for the (axial)-vector mesons. Furthermore, all couplings in the second line are defined to be dimensionless. The subindex of the operators stands for the kind of vertex described, i.e., ${\mathcal O}^i_{VPPP}$ gives a coupling between one Vector and three Pseudoscalars. For the explicit form of the operators in the last line, see \cite{vap}, \cite{vvp}, \cite{paper}.\\
\noindent Symmetries of low-energy QCD are not enough for R$\chi$T to describe it up to the perturbative region. It is mandatory to match order parameters of spontaneous chiral symmetry breaking with partonic QCD related quantities. The matching of $2$ and $3$-point Green Functions in the OPE of QCD and in R$\chi$T has been shown to be a fruitful procedure \cite{rcht2}, \cite{vap}, \cite{vvp}, \cite{amo}, \cite{knecht},  \cite{consistent}. Additionally, we will demand to the vector and axial-vector form factors a Brodsky-Lepage-like behaviour \cite{brodskylepage}.\\
\noindent While symmetry fully determines the structure of the operators, it is the QCD-ruled short-distance behaviour who restricts certain combinations of couplings rendering R$\chi$T predictive provided we fix these few remaining parameters restoring to phenomenology
. The set of couplings determined by this procedure depends on the particular process under study. In particular some couplings that were fitted phenomenologically in $\rho^0\to e^+e^-,\pi^+\pi^-$ and  $a_1\to \pi\gamma,\rho\pi$ are predicted now
.
\section{ASYMPTOTIC BEHAVIOUR AND QCD CONSTRAINTS} \label{Asymptotic_behaviour}
\noindent There are 24 unknown couplings in $\mathcal{L}_{R\chi T}$ (\ref{Full_Lagrangian}) that may appear in the calculation of three meson decays of the $\t$.  They are reduced when computing the Feynman diagrams involved. There only appear $F_V$, $G_V$, three combinations of the $\lbrace \lambda_i\rbrace_{i=1}^5$, four of the $\lbrace c_i\rbrace_{i=1}^7$, two of the $\lbrace d_i\rbrace_{i=1}^4$ and four of the $\lbrace g_i\rbrace_{i=1}^5$. The number of free parameters has been reduced from 24 to 15.\\
\noindent We require the form factors of the $\mathcal{A}^\m$ and $\mathcal{V}^\m$ currents into $KK\p$ modes vanish at infinite transfer of momentum. As a result, we obtain constraints \cite{paper}, \cite{Roig:2007yp} among all axial-vector current couplings but $\lambda_0$, that are also the most general ones satisfying the demanded asymptotic behaviour in $\t\to3\p\n_\t$, studied along these lines in \cite{t3p}. Proceeding analogously with the vector current form factor results in five additional restrictions \cite{paper}. From the 24 initially free couplings in Eq. (\ref{Full_Lagrangian}), only five remain free: $c_4$, $c_1\,+\,c_2\,+\,8\,c_3\,-\,c_5$, $d_1\,+\,8\,d_2\,-\,d_3$, $g_4$ and $g_5$. After fitting $\Gamma(\omega\to3\pi)$ -using some of the relations in Ref. \cite{vvp}- only $c_4$ and $g_4$ remain unknown. We obtained them from BaBar data on $e^+e^-\to KK\pi$
.
\section{RESULTS AND CONCLUSIONS} \label{results_and_conclusions}
\begin{figure}[h]
The plot presented here corresponds to:
\begin{eqnarray} \label{coups_final}
M_{a_1} = 1.17 \,\,\mathrm{GeV},\,\,c_4 = -0.04,\,\,g_4=-0.5.
\end{eqnarray}
\begin{center}
\includegraphics[scale=0.3,angle=-90]{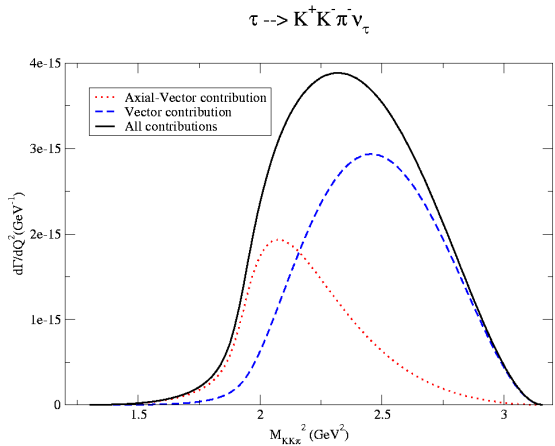}
\end{center}

\caption{\small{Spectral function of $\tau\to K^+K^-\pi^-\nu_\tau$ obtained with the set of parameters (\ref{coups_final}).}}
\label{K+K-pi-}
\end{figure}

\noindent Relating the measured isovector component of $e^+e^-\to KK\pi$ to the total isovector cross-section, as done in Ref.~\cite{aleph}, is not trivial at all~\cite{Roig:2008ev}. The issue is discussed in detail in our article \cite{paper}. We conclude that the measured isovector component in $e^+e^-\to K_S K^{\pm} \pi^{\mp}$ does not provide the total isovector component for the (partially) inclusive decay $e^+e^-\to KK\pi$. Once some assumptions are used, one can employ $CVC$ to relate $e^+e^-\to KK\pi$ to $\t \to KK\pi\nu_\t$ 
. We fitted our expressions to BaBar data \cite{babar2} obtaining $c_4=-0.044\pm 0.005$. Our results for the decay widths of the considered channels are consistent with the PDG values \cite{PDG2006}. To confront our predictions for the spectra with the forthcoming experimental data is a necessary task.\\
We have obtained several sets of allowed values for those quantities that were still free after the high-energy conditions were imposed \cite{paper}.
We get two predictions:
\begin{itemize}
\item The ratio of the vector current to all contributions is, for all charge modes:
\begin{equation}
\frac{\Gamma_V\left(\tau\to KK\pi\n_\t\right)}{\Gamma\left(\tau\to KK\pi\n_\t\right)} \sim 0.5.
\end{equation}
\item The spectra of the two independent $KK\pi$ decay modes of the $\tau$. We show in Figure \ref{K+K-pi-} that one for $\tau\to K^+K^-\pi^-\n_\t$. The shape of the spectral function is similar for $\tau\to K^-K^0\pi^0\n_\t$. In both cases, the axial-vector current dominates the low-energy region, while the vector one peaks at higher values of $Q^2$.
\end{itemize}


We conclude by emphasizing that the phenomenological approach that we follow includes every relevant (and known) piece of QCD: We preserve the correct chiral limit at low energies that gives the right normalization to our form factors, we employ large-$N_C$ QCD arguments to use our theory in terms of mesons and it fulfills the high-energy conditions of the fundamental theory at the mesonic level. While we remain predictive by including only the lightest multiplet of axial-vector and vector resonances, our Lagrangian can be enlarged in a systematic way to account for the contribution of higher resonance states, whose parameters may be fitted to experiment.\\
This framework can easily be extended to study exclusive hadronic channels in $e^+e^-$ collisions, as we plan to do. Once the spectra for $\tau\to KK\pi\nu_\tau$ are presented, our work may shed some light on the issue of the importance of isospin violating corrections when relating tau and $e^+e^-$ hadronic cross sections.\\
Our study can be very useful for the B-factories BaBar and Belle, that are already analyzing huge samples with good quality data, the tau-charm factory BES-III and a future super-$B$ factory; specially, if our expressions are implemented in the currently used libraries for both types of processes, namely TAUOLA~\cite{TAUOLA} and PHOKHARA~\cite{PHOKHARA}.

\end{document}